\documentclass[graybox]{svmult}

\usepackage{mathptmx}       
\usepackage{helvet}         
\usepackage{courier}        
\usepackage{type1cm}        
\usepackage{makeidx}         
\usepackage{graphicx}        
\usepackage{multicol}        
\usepackage[bottom]{footmisc}
\usepackage{url}

\usepackage{ifthen}
\usepackage{makeglos}

\graphicspath{{./}{./graphics/}{./graphics/eps/}{./graphics/pdf/}}
\usepackage{amssymb}
\usepackage{watermark}
\usepackage{amsfonts}
\usepackage{epsfig}
\usepackage{lettrine}
\usepackage{rotating}
\usepackage{multirow}
\usepackage{color}
\usepackage{colortbl}
\usepackage{subfigure}
\usepackage{rotating}
\usepackage{pdflscape}
\usepackage{afterpage}
\usepackage{capt-of}
\usepackage{lipsum}

\newcommand{\buildbook}{false}

\makeindex             

\makeglossary

\begin{document}

%
%
%

\ifthenelse{\equal{true}{\buildbook}}{
\title{Introduction to Presentation Attack Detection \\in Fingerprint Biometrics}
}
{
\title*{Introduction to Presentation Attack Detection \\in Fingerprint Biometrics}
}


\author{Javier Galbally, Julian Fierrez, Raffaele Cappelli, and Gian Luca Marcialis}


\institute{Javier Galbally \at European Commission, Joint Research Centre, Ispra, Italy, \email{javier.galbally@ec.europa.eu}
\and Julian Fierrez \at Biometrics and Data Pattern Analytics - BiDA Lab, Universidad Autonoma de Madrid, Madrid, Spain, \email{julian.fierrez@uam.es}
\and Raffaele Capelli \at Università di Bologna, Cesena, Italy, \email{raffaele.cappelli@unibo.it}
\and Gian Luca Marcialis \at Università di Cagliari, Italy, \email{marcialis@unica.it}}

%

%
\maketitle

\abstract{
This chapter provides an introduction to Presentation Attack Detection (PAD) in
fingerprint biometrics, also coined anti-spoofing, describes early developments in this field, and briefly summarizes recent trends and open issues.
}

\section{Introduction}
\label{sec:intro}

``\emph{Fingerprints cannot lie, but liars can make fingerprints}''.
Unfortunately, this paraphrase of an old quote attributed to Mark
Twain\footnote{Figures do not lie, but liars do figure.} has been
proven right in many occasions now.

As the deployment of fingerprint\index{Fingerprint} systems keeps growing year after
year in such different environments as airports, laptops or mobile
phones, people are also becoming more familiar to their use in
everyday life and, as a result, the security weaknesses of
fingerprint sensors are becoming better known to the general public.
Nowadays it is not difficult to find websites or even tutorial
videos, which give detail guidance on how to create fake
fingerprints which may be used for spoofing biometric systems.

As a consequence, the fingerprint stands out as one of the biometric
traits which has arisen the most attention not only from researchers
and vendors, but also from the media and users, regarding its
vulnerabilities to Presentation Attacks (PAs)\glossary{PA: Presentation Attack}\index{Presentation Attack (PA)}
(aka spoofing\glossary{Spoofing (term to be deprecated): attempt to impersonate a biometric system, \gsee{PA}}\index{Spoofing (term to be deprecated)}). 
This increasing interest of the
biometric community in the security evaluation of fingerprint
recognition systems against presentation attacks has lead to the
creation of numerous and very diverse initiatives in this field: the
publication of many research works disclosing and evaluating
different fingerprint presentation attack approaches
\cite{putte00fakeFingerprints,matsumoto02fakeFingerprints,thalheim02directAttacks,busch14fingerPAD};
the proposal of new Presentation Attack Detection (PAD)\glossary{PAD: Presentation Attack Detection}\index{Presentation Attack Detection (PAD)}
(aka anti-spoofing\glossary{Anti-Spoofing (term to be deprecated): countermeasure to spoofing, \gsee{PAD}}
\index{Anti-Spoofing (term to be deprecated)}) methods
\cite{schuckers03LivenessSweat,antonelli06LivenessDistortion,galbally12LDfps};
related book chapters
\cite{franco08ChapterAntispoofingFPs,zli09enciclopedia}; PhD and
MsC Thesis which propose and analyse different fingerprint PA
and PAD techniques
\cite{coli08LDfpsThesis,sandstrom04LDspoofingMsC,lane05SpoofFPsMsC,blomme03SpoofingFps};
several patented fingerprint PAD mechanisms both for
touch-based and contactless systems
\cite{lapsley98LDfpsBloodFlowPatent,setlak99LDfpsPatent,kallo01LDfpsPatent,diaz08fpsLDpatent,kim11TouchlessLDpatent};
the publication of Supporting Documents and Protection Profiles in
the framework of the security evaluation standard Common Criteria
for the objective assessment of fingerprint-based commercial systems
\cite{SD-CC11CAFVM,PP-CC08BSIspoofingFPs}; the organization of
competitions focused on vulnerability assessment to fingerprint presentation attacks
\cite{ghiani17livdetreview,casula21livdet,orru19livdet}
; the acquisition of specific
datasets for the evaluation of fingerprint protection methods
against direct attacks
\cite{galbally11directAttacksTS,abhyankar09LDfpsPerspirationWavelets,spinoulas21multidata},
the creation of groups and laboratories which have the evaluation of
fingerprint security as one of their major tasks
\cite{BVAEG,NPL,BWG}; or the acceptance of several European
Projects on fingerprint PAD as one of their main
research interests \cite{BEAT,TabulaRasa}.


The aforementioned initiatives and other analogue studies, have
shown the importance given by all parties involved in the
development of fingerprint-based biometrics to the improvement of
the systems security and the necessity to propose and develop
specific protection methods against PAs in order to
bring this rapidly emerging technology into practical use. This way,
researchers have focused on the design of specific countermeasures
that enable fingerprint recognition systems to detect fake samples
and reject them, improving this way the robustness of the
applications.


In the fingerprint field, besides other PAD approaches
such as the use of multibiometrics or challenge-response methods,
special attention has been paid by researchers and industry to the
so-called \emph{liveness detection} techniques. These algorithms use
different physiological properties to distinguish between real and
fake traits. Liveness assessment methods represent a challenging
engineering problem as they have to satisfy certain demanding
requirements \cite{maltoni09book}: $(i)$ non-invasive, the
technique should in no case be harmful for the individual or require
an excessive contact with the user; $(ii)$ user friendly, people
should not be reluctant to use it; $(iii)$ fast, results have to be
produced in a very reduced interval as the user cannot be asked to
interact with the sensor for a long period of time; $(iv)$ low cost,
a wide use cannot be expected if the cost is excessively high; $(v)$
performance, in addition to having a good fake detection rate, the
protection scheme should not degrade the recognition performance
(i.e., false rejection) of the biometric system.

Liveness detection methods are usually classified into one of two
groups: $(i)$ \emph{Hardware-based} techniques, which add some
specific device to the sensor in order to detect particular
properties of a living trait (e.g., fingerprint sweat, blood
pressure, or odor); $(ii)$ \emph{Software-based} techniques, in this
case the fake trait is detected once the sample has been acquired
with a standard sensor (i.e., features used to distinguish between
real and fake traits are extracted from the biometric sample, and
not from the trait itself).

The two types of methods present certain advantages and drawbacks
over the other and, in general, a combination of both would be the
most desirable protection approach to increase the security of
biometric systems. As a coarse comparison, hardware-based schemes
usually present a higher fake detection rate, while software-based
techniques are in general less expensive (as no extra device is
needed), and less intrusive since their implementation is
transparent to the user. Furthermore, as they operate directly on
the acquired sample (and not on the biometric trait itself),
software-based techniques may be embedded in the feature extractor
module which makes them potentially capable of detecting other types
of illegal break-in attempts not necessarily classified as presentation
attacks. For instance, software-based methods can protect the system
against the injection of reconstructed or synthetic samples into the
communication channel between the sensor and the feature extractor
\cite{Cappelli07PAMIReconstruction,cappelli09SfingeChapter}.

Although, as shown above, a great amount of work has been done in
the field of fingerprint PAD and big advances have
been reached over the last two decades, the attacking methodologies have
also evolved and become more and more sophisticated. This way, while
many commercial fingerprint readers claim to have some degree of
PAD embedded, many of them are still vulnerable to
presentation attack attempts using different artificial fingerprint samples.
Therefore, there are still big challenges to be faced in the
detection of fingerprint direct attacks.\footnote{\url{https://www.iarpa.gov/index.php/research-programs/odin/}}

This chapter represents an introduction to the problem of fingerprint PAD \cite{galbally14TIP,hadid15SPMspoofing}. More comprehensive and up to date surveys of recent advances can be
found elsewhere \cite{ross14surveyPAD,Busch14surveyPAD,kram21surveyPAD,ragha21surveyPAD}.
The rest of the chapter is structured as follows. An overview into early works in the field of fingerprint
PAD is given is Section~\ref{sec:sota}, while Section~\ref{sec:updsota} provides a summary of recent trends and main open issues. A brief
description of large and publicly available fingerprint spoofing databases is presented in Section~\ref{sec:dbs}. Conclusions are finally drawn in
Sect.~\ref{sec:conclusions}.

\section{Early Works in Fingerprint Presentation Attack
Detection}
\label{sec:sota}


The history of fingerprint forgery in the forensic field is probably
almost as old as that of fingerprint development and classification
itself. In fact, the question of whether or not fingerprints could
be forged was positively answered \cite{wehde24forgingFPs} several
years before it was officially posed in a research publication
\cite{water36FpsForgery}.


Regarding modern automatic fingerprint recognition systems, although
other types of attacks with dead \cite{sengottuvelan07LiveDeadFPs}
or altered \cite{yoon12AlteredFPs} fingers have been reported,
almost all the available vulnerability studies regarding presentations
attacks are carried out either by taking advantage of the residual
fingerprint left behind on the sensor surface, or by using some type
of gummy fingertip (or even complete prosthetic fingers)
manufactured with different materials (e.g., silicone, gelatin,
plastic, clay, dental molding material or glycerin). In general,
these fake fingerprints may be generated with the cooperation of the
user, from a latent fingerprint or even from a fingerprint image
reconstructed from the original minutiae template
\cite{willis98SpoofFPs,putte00fakeFingerprints,matsumoto02fakeFingerprints,thalheim02directAttacks,sten03SpoofFpsPrecise100SC,wiehe04FpsSpoofing,galbally09FPsPRLda,barral09spoofingFpsGlycerin,galbally11directAttacksTS}.

These very valuable works and other analogue studies, have
highlighted the necessity to develop efficient protection methods
against presentation attacks. One of the first efforts in fingerprint
PAD initiated a research line based on the analysis of the
skin perspiration pattern which is very difficult to be faked in an
artificial finger
\cite{schuckers03LivenessSweat,parthasaradhi05fpsLDperspiration}.
These pioneer studies, which considered the periodicity of sweat and
the sweat diffusion pattern, were later extended and improved in two
successive works applying a wavelet-based algorithm and adding
intensity-based perspiration features
\cite{schuckers04LivenessWavelet,schuckers06LivenessSPIE}. These
techniques were finally consolidated and strictly validated on a
large database of real, fake and dead fingerprints acquired under
different conditions in
\cite{abhyankar09LDfpsPerspirationWavelets}. Recently, a novel
region-based liveness detection approach also based on perspiration
parameters and another technique analyzing the valley noise have
been proposed by the same group
\cite{tan08FPsLDnoise,decann09FPsLDperspiration}. Part of these
approaches have been implemented in commercial products
\cite{NexIDBiometrics}, and have also been combined with other
morphological features
\cite{abhyankar06LDfpsMultiresolutionTexture,marasco10LDfpsMultiTexture}
in order to improve the presentation attack detection rates
\cite{marasco12fpsLDperspiration}.

A second group of fingerprint liveness detection techniques has
appeared as an application of the different fingerprint distortion
models described in the literature
\cite{Cappelli01Distortion,bazen03FPsDistortion,chen05FPsDeformation}.
These models have led to the development of a number of liveness
detection techniques based on the flexibility properties of the skin
\cite{jain05LivenessDeformation,antonelli06LivenessDistortion,zhang07FPsDistortion,yau07fpsLDcolor}.
In most of these works the user is required to move his finger while
pressing it against the scanner surface, thus deliberately
exaggerating the skin distortion. When a real finger moves on a
scanner surface, it produces a significant amount of distortion,
which can be observed to be quite different from that produced by
fake fingers which are usually more rigid than skin. Even if highly
elastic materials are used, it seems very difficult to precisely
emulate the specific way a real finger is distorted, because the
behavior is related to the way the external skin is anchored to the
underlying derma and influenced by the position and shape of the
finger bone.

Other liveness detection approaches for fake fingerprint detection
include: the combination of both perspiration and elasticity related
features in fingerprint image sequences \cite{jia07FPsLDdefPersp};
fingerprint-specific quality-related features
\cite{uchida04LDfpsQuality,galbally12LDfps}; the combination of the
local ridge frequency with other multiresolution texture parameters
\cite{abhyankar06LDfpsMultiresolutionTexture}; techniques which,
following the perspiration-related trend, analyze the skin sweat
pores visible in high definition images
\cite{marcialis10fpsLDpores,memon11LDfpsPores}; the use of electric
properties of the skin \cite{martinsen07FPsLDelectric}; using
several image processing tools for the analysis of the finger tip
surface texture such as wavelets \cite{moon05FPsLDwavelet}, or
three very related works using gabor filters
\cite{nikam09LDfpsGaborJournal}, ridgelets
\cite{nikam09fpsLDridgelet} and curvelets
\cite{nikam10LDfpsCurveletJournal}; analyzing different
characteristics of the Fourier spectrum of real and fake fingerprint
images
\cite{coli07FPsLDpower,jin07FPsLDfourier,jin10LDfpsFourier,lee09LDfpsFT,marcialis12LDfpsFakeCharacteristics}.

A critical review of some of these solutions for fingerprint
liveness detection was presented in \cite{coli07LDfpsSurvey}. In a
subsequent work \cite{roli08LivenessComparative}, the same authors
gave a comparative analysis of the PAD methods efficiency.
In this last work we can find an estimation of some of the best
performing static (i.e., measured on one image) and dynamic (i.e.,
measured on a sequence of images) features for liveness detection,
that were later used together with some fake-finger specific
features in \cite{marcialis12LDfpsFakeCharacteristics} with very
good results. Different static features are also combined in
\cite{choi09LDfpsMultipleStatic}, significantly improving the
results of the individual parameters. Other comparative results of
different fingerprint PAD techniques are available in the
results of the 2009 and 2011 Fingerprint Liveness Detection
Competitions (LivDet 2009 and LivDet 2011)
\cite{marcialis09livdet,livdet11}.

In addition, some very interesting hardware-based solutions have
been proposed in the literature applying: multispectral imaging
\cite{nixon05fpsLDmultispectral,rowe08LDfpsMultispectralChapter},
an electrotactile sensor \cite{yau08fpsLDelectrotactile}, pulse
oxiometry \cite{reddy08LDfpsOximetryJournal}, detection of the
blood flow \cite{lapsley98LDfpsBloodFlowPatent}, odor detection
using a chemical sensor \cite{baldiserra06LivenessOdor}, or a
currently very active research trend based on Near Infrared (NIR)
illumination and Optical Coherence Tomography (OCT)
\cite{cheng06LDfpsOCTjournal,manapuram06LDfpsOCTjournal,cheng07LDfpsOCTjournal,larin08LDfpsOCTconf,chang11LDfpsNIRchapter,nasiri11LDfpsOCTjournal}.

More recently, a third type of protection methods which fall out of
the traditional two-type classification software- and hardware-based
approaches have been started to be analyzed in the field of
fingerprint PAD. These protection techniques focus on the
study of biometric systems under direct attacks at the \emph{score
level}, in order to propose and build more robust matchers and
fusion strategies that increase the resistance of the systems
against presentation attack attempts
\cite{rattani12AntispoofScoreLevel,marasco12LDfusionScores,hariri11AntispoofingMultimodalFusion,akhtar11AntispoofingMultimodalFusion,akhtar12spoofingMultimodal}.

Outside the research community, some companies have also proposed
different methods for fingerprint liveness detection such as the
ones based on ultrasounds \cite{Ultrascan,Optel}, light
measurements \cite{posid}, or a patented combination of different
unimodal experts \cite{VirdiTech}. A comparative study of the
PAD capabilities of different commercial fingerprint
sensors may be found in \cite{kang03testliveness}.

Although the vast majority of the efforts dedicated by the biometric
community in the field of fingerprint presentation attacks and PAD are
focused on touch-based systems, some works have also
been conducted to study the vulnerabilities of contactless
fingerprint systems against direct attacks and some protection
methods to enhance their security level have been proposed
\cite{parthasaradhi05fpsLDperspiration,diaz08fpsLDpatent,wang09touchlessLD}.

The approaches mentioned above represent the main historical developments in
fingerprint PAD until ca. 2012–2013. For a survey of more recent and advanced
methods in the last 10 years we refer the reader to \cite{ross14surveyPAD,Busch14surveyPAD,kram21surveyPAD,ragha21surveyPAD}, and the ODIN program.\footnote{\url{https://www.iarpa.gov/index.php/research-programs/odin/}}

\section{A brief view on where we are}
\label{sec:updsota}
In the next chapters of the book, the reader will be able to find information about the most recent advances in fingerprint presentation attack detection. This section merely summarizes some ongoing trends in the development of PADs and some of the main open issues.

As stated in the previous Section, independent and general-purpose descriptors were proposed for feature extraction since from 2013 \cite{Busch14surveyPAD}. In general, these feature looked for minute details of the fake image which are added or deleted, impossible to catch by the human eye. This was typically done by appropriate banks of filters aimed at derive a set of possible patterns. The related feature sets can be adopted to distinguish live from fake fingerprint by machine learning methods. 

``Textural features'' above looked as the most promising until the advent of deep learning approaches \cite{kram21surveyPAD,ragha21surveyPAD}. These, thanks to the increased availability of data sets, allowed the design of a novel generation of fingerprint PAD \cite{spinoulas21multidata,chugh18spoofbuster,park19tinyconvolutional} which exploited the concept of ``patch'', a very small portion of the fingerprint image to be processed instead of taking the image as a whole input to the network. However, textural features have not yet left behind because of their expressive power and the fact that they explicitly rely on the patch definition \cite{xia20weber,agarwal21comparative}.

Among the main challenges to be faced with in the near future, it is important to mention \cite{jain2021biometricsTbV}:
\begin{itemize}

\item assessing the robustness of anti-spoofing methods against novel presentation attacks in terms of fabrication strategy, adopted materials, sensor technology; for instance, in \cite{chugh21spoofdetector} it has been shown that the PAD error rates of software-based approaches can show a three-fold increase when tested on PA materials not seen during training;

\item designing effective methods to embed PAD in fingerprint verification systems \cite{micheletto21embedded}, including the need for computationally efficient PAD techniques, to be used on low-resources systems such as embedded devices and low-cost smartphones;

\item improving explainability of PAD systems; the use of CNNs is providing great benefits to fingerprint PAD performance, but such solutions are usually considered as ''black boxes'' shedding little light on how and why they actually work. It is important to gain insights into the features that CNNs learn, so that system designers and maintainers can understand why a decision is made and tune the system parameters if needed.

\end{itemize}

\section{Fingerprint Spoofing Databases}
\label{sec:dbs}


The availability of public datasets comprising real and fake
fingerprint samples and of associated common evaluation protocols is
basic for the development and improvement of fingerprint
PAD methods.

However, in spite of the large amount of works addressing the
challenging problem of fingerprint protection against direct attacks
(as shown in Section~\ref{sec:sota}), in the great majority of them,
experiments are carried out on proprietary databases which are not
distributed to the research community.

Currently, the two largest fingerprint spoofing databases publicly
available for researchers to test their PAD algorithms
are:

\begin{itemize}

\item LivDet DBs (LivDet 2009-2021 DBs) 
\cite{ghiani17livdetreview,casula21livdet,orru19livdet}: These data sets, which share the acquisition protocols and part of the samples, are available from 2009 to 2021 Fingerprint Liveness Detection Competitions
websites\footnote{\url{http://livdet.diee.unica.it}}\footnote{\url{http://people.clarkson.edu/projects/biosal/fingerprint/index.php}}
and are divided into the same train and test set used in the
official evaluations. Over seven editions, LivDet shared with the research community over 20,000 fake fingerprint images made up of a large set of materials (play doh, silicone, gelatine, latex...) on a wide brands of optical and solid-state sensors. Over years, LivDet competitions also proposed challenges as the evaluation of embedding fingerprint PAD and matching \cite{casula21livdet,orru19livdet}, and of a novel approach to provide spoofs called ``Screenspoof'' directly from the user's smartphone screen \cite{casula21livdet}. The LivDet datasets are available for researchers by signing the license agreement.

\item ATVS-Fake Fingerprint DB (ATVS-FFp DB) \cite{galbally11directAttacksTS}: This database is
available from the Biometrics group at UAM\footnote{\url{http://biometrics.eps.uam.es/}}. It contains over
3,000 real and fake fingerprint samples coming from 68 different
fingers acquired using a flat optical sensor, a flat capacitive
sensor and a thermal sweeping sensor. The gummy fingers were
generated with and without the cooperation of the user (i.e.,
recovered from a latent fingerprint) using modeling silicone.

\end{itemize}

\section{Conclusions}
\label{sec:conclusions}

The study of the vulnerabilities of biometric systems against
presentation attacks has been a very active field of research in recent
years \cite{nixon08AntispoofHandbook}. This interest has lead to big
advances in the field of security-enhancing technologies for
fingerprint-based applications. However, in spite of this noticeable
improvement, the development of efficient protection methods against
known threats (usually based on some type of self-manufactured gummy
finger) has proven to be a challenging task.

Simple visual inspection of an image of a real fingerprint and its
corresponding fake sample shows that the two images can be very
similar and even the human eye may find it difficult to make a
distinction between them after a short inspection. Yet, some
disparities between the real and fake images may become evident once
the images are translated into a proper feature space. These
differences come from the fact that fingerprints, as 3-D objects,
have their own optical qualities (absorption, reflection,
scattering, refraction), which other materials (silicone, gelatin,
glycerin) or synthetically produced samples do not possess.
Furthermore, fingerprint acquisition devices are designed to provide
good quality samples when they interact, in a normal operation
environment, with a real 3-D trait. If this scenario is changed, or
if the trait presented to the scanner is an unexpected fake
artifact, the characteristics of the captured image may
significantly vary.

In this context, it is reasonable to assume that the image quality
properties of real accesses and fraudulent attacks will be
different and therefore image-based presentation attack detection in fingerprint biometrics would be feasible. Key early works in this regard have been summarized in the present chapter.

Overall, the chapter provided a general overview of the
progress which was initially made in the field of
fingerprint PAD and a brief summary about current achievements, trends, and open issues, which will be further developed in the next chapters.

\begin{acknowledgement}
This work was mostly done (2nd Edition of the book) in the context of the
TABULA RASA\glossary{TABULA RASA: Trusted Biometrics under Spoofing Attacks.} and
BEAT\glossary{BEAT: Biometrics Evaluation and Testing.}
projects funded under the 7th Framework Programme of EU\index{7th European Framework Programme (FP7)}. The 3rd Edition update has been made in the context of EU H2020 projects PRIMA and TRESPASS-ETN. This work was also partially supported by the Spanish project BIBECA (RTI2018-101248-B-I00 MINECO/FEDER).\end{acknowledgement}

\ifthenelse{\equal{false}{\buildbook}}{
\printindex
\printglossary
\bibliographystyle{spmpsci}
\bibliography{references}
}

\end{document}